\documentclass{article}
\usepackage{amsmath,amssymb}
\usepackage{fullpage}
\usepackage[utf8]{inputenc}

\usepackage{amsmath,amsfonts,amssymb,amsthm}
\usepackage[colorlinks=true,linkcolor=red,citecolor=blue,urlcolor=red]{hyperref}
\usepackage{paralist}
\usepackage{framed}
\usepackage{verbatim}
\usepackage{cleveref}
\usepackage[dvipsnames]{xcolor}

\usepackage{mathptmx}
\usepackage[text={6.45in,8.95in}]{geometry}

\newenvironment{reminder}[1]{\smallskip
	\noindent {\bf Reminder of #1.  }\em}{\smallskip}

\newtheorem{theorem}{Theorem}[section]
\newtheorem{hypothesis}{Hypothesis}

\newtheorem{proposition}{Proposition}[section]
\newtheorem{corollary}{Corollary}[section]
\newtheorem{lemma}{Lemma}[section]

\def \eps{{\varepsilon}}
\def\poly{\text{poly}}

\def \TC {{\sf TC}}

\def \ACC {{\sf ACC}}
\def \AC {{\sf AC}}
\def \CC {{\sf CC}}
\def \Z {{\mathbb Z}}

\def \F {{\mathbb F}}

\def \NC {{\sf NC}}

\def \PP {{\sf PP}}

\def \SYM {\text{SYM}}
\def \AND {\text{AND}}
\def \MOD {\text{MOD}}

\def\ShowAuthNotes{1}
\ifnum\ShowAuthNotes=1
\newcommand{\authnote}[2]{\ \\ \textcolor{red}{\parbox{0.9\linewidth}{[{\footnotesize {\bf #1:} { {#2}}}]}}\newline}
\else
\newcommand{\authnote}[2]{}
\fi

\usepackage{enumitem}

\title{Smaller ACC0 Circuits for Symmetric Functions\thanks{Supported by NSF CCF-1909429, NSF CCF-1741615, and a Frank Quick Faculty Research Innovation Fellowship.}}
\author{Brynmor Chapman\\MIT 
\and Ryan Williams\thanks{This work was done while the author was visiting the Simons Institute for the Theory of Computing, participating in the \emph{Theoretical Foundations of Computer Systems} and \emph{Satisfiability: Theory, Practice, and Beyond} programs.} \\MIT}
\date{ } 

\begin{document}

\maketitle

\begin{abstract} What is the power of 
constant-depth circuits with $\text{MOD}_m$ gates, that can count modulo $m$? Can they efficiently compute MAJORITY and other symmetric functions?
When $m$ is a constant prime power, the answer is well understood. In this regime, Razborov and Smolensky proved in the 1980s that MAJORITY and $\text{MOD}_m$ require super-polynomial-size $\text{MOD}_q$ circuits, where $q$ is any prime power not dividing $m$.
However, relatively little is known about the power of $\text{MOD}_m$ gates when $m$ is not a prime power. For example, it is still open whether every problem decidable in exponential time can be computed by depth-$3$ circuits of polynomial-size and only $\text{MOD}_6$ gates.

In this paper, we shed some light on the difficulty of proving lower bounds for $\text{MOD}_m$ circuits, by giving new upper bounds. We show how to construct $\text{MOD}_m$ circuits computing symmetric functions with non-prime power $m$, with size-depth tradeoffs that beat the longstanding lower bounds for $\AC^0[m]$ circuits when $m$ is a prime power. Furthermore, we observe that our size-depth tradeoff circuits have essentially optimal dependence on $m$ and $d$ in the exponent, under a natural circuit complexity hypothesis. 

For example, we show that for every $\eps > 0$, every symmetric function can be computed using $\text{MOD}_m$ circuits of depth $3$ and $2^{n^{\eps}}$ size, for a constant $m$ depending only on $\eps > 0$. In other words, depth-$3$ $\CC^0$ circuits can compute any symmetric function in \emph{subexponential} size. This demonstrates a significant difference in the power of depth-$3$ $\CC^0$ circuits, compared to other models: for certain symmetric functions, depth-$3$ $\AC^0$ circuits require $2^{\Omega(\sqrt{n})}$ size [H{\aa}stad 1986], and depth-$3$ $\AC^0[p^k]$ circuits (for fixed prime power $p^k$) require
$2^{\Omega(n^{1/6})}$ size [Smolensky 1987]. Even for depth-two $\MOD_p \circ \MOD_m$ circuits, $2^{\Omega(n)}$ lower bounds were known [Barrington Straubing Th{\'e}rien 1990].

\end{abstract}

\thispagestyle{empty}
\setcounter{page}{0}
\newpage

\section{Introduction}

We consider constant-depth circuits in which every (unbounded fan-in) gate (called a $\MOD_m$ gate) determines whether the sum of its inputs is divisible by a small constant integer $m$. Although the model looks rather peculiar, constant-depth circuits with constant moduli gates (a.k.a.\ $\CC^0$ circuits, a.k.a.\ pure-$\ACC$ circuits~\cite{Yao90}) have been a longstanding and fundamental roadblock in the way of improved circuit complexity lower bounds. Since their identification over 30 years ago~\cite{Barrington89,BarringtonT88}, scant progress has been made on lower bounds against $\CC^0$ circuits, and their close cousin $\ACC^0$ which includes AND and OR in the gate basis. Some exceptions include work focusing on special cases of the problem~(e.g.,~\cite{BarringtonBR94,GrolmuszTardos,ChattopadhyayGPT06,ChattopadhyayW09}), uniform lower bounds~\cite{AllenderG94}, and work proving strong lower bounds but only for functions whose complexity is in $\mathsf{QuasiNP}$ or higher~(e.g.,~\cite{WilliamsJACM14,ChenP19,MurrayW20,ChenLW20}). If there has ever been a ``circuit complexity winter'', $\CC^0$ circuits are at least partly to blame.

Besides our own ignorance, could there be deeper reasons why $\CC^0$ circuits have been so difficult for showing limitations? In this paper, we explore the possibility that $\CC^0$ circuits may be powerful, focusing on the natural class of \emph{symmetric Boolean functions} whose output depends only on the number of ones in the input. Although it has been conjectured for many years that the AND function does not have polynomial-size $\CC^0$ circuits~(\cite{Barrington-Immerman-Straubing90,Therien94,StraubingT06})\footnote{See~\cite{HansenK10} for an interesting counterpoint. They show that \emph{probabilistic} $\CC^0$ circuits can in fact compute AND efficiently, and that the $\text{AND} \in \CC^0$ problem is equivalent to a certain $\CC^0$ derandomization question.} our results show that low-depth $\MOD_m$ circuits with arbitrary but fixed modulus $m$ can actually compute arbitrary symmetric Boolean functions (such as MAJORITY) much more efficiently than low-depth circuits with AND, OR, and $\MOD_q$ gates, when $q$ is a prime power.

It is well-known that $\AC^0$ circuits, which consist of AND, OR, NOT gates and have constant-depth, require  $\exp(\Omega(n^{1/(d-1)}))$ size to compute arbitrary symmetric functions in depth $d$~\cite{Hastad86}. In recent work, Oliveira, Santhanam, and Srinivasan~\cite{OliveiraSS19} have shown that PARITY gates (a.k.a.\ $\MOD_2$ gates) can help compute symmetric functions more efficiently than what AND, OR, NOT can accomplish in constant depth. In particular, they show that $\AC^0[2]$ circuits (with AND, OR, and PARITY) of depth $4$ can compute MAJORITY in $\exp(\Theta(n^{1/4}))$ size, depth $d \geq 5$ can compute symmetric functions in size $\exp({\widetilde{O} (n^{\frac{2}{3(d - 4)}} )})$, and they show a size lower bound of $\exp(\Omega(n^{1/(2d-4)}))$ for the MAJORITY function, improving~\cite{Razborov87,Smolensky87}.

\paragraph{Smaller $\text{MOD}_m$ Circuits.} Could even smaller circuits for symmetric functions be achieved  using $\MOD_m$ gates, for other composite $m$? It turns out that this is possible. In fact, even in depth three, any symmetric function can computed with a $\text{MOD}_m$ circuit of size $2^{n^{\eps}}$ for any desired $\eps > 0$.

\begin{theorem}\label{thm:depth-3-subexp}
For every $\eps > 0$, there is a modulus $m \le (1/\eps)^{2/\eps}$ such that every symmetric function on $n$ bits can be computed by depth-$3$ $\MOD_m$ circuits of $\exp(O(n^{\eps}))$ size. In fact, the circuits have the form $\MOD_{p_1} \circ \MOD_{p_2 \cdots p_r} \circ \MOD_{p_1}$, where $p_1,\ldots,p_r$ are distinct primes.\footnote{The $G \circ H \circ I$ notation means that the output gate has type $G$, on the middle layer there are gates only of type $H$, and on the bottom layer (nearest the inputs) there are only gates of type $I$.}
\end{theorem}

That is, without any AND/OR gates, we can obtain $\CC^0$ circuits that beat the longstanding lower bounds for $\AC^0[q]$ circuits, for prime power $q$.

It has been known for decades~\cite{BarringtonST90} that \emph{depth-two} $\MOD_{p} \circ \MOD_{m}$ circuits (and $\CC^0[p] \circ \MOD_m$ circuits) require $2^{\Omega(n)}$ size to compute the AND function, where $p$ is a prime and $m$ is an arbitrary composite, and that only certain restricted symmetric functions could be computed in subexponential-size and depth-two~\cite{GrolmuszTardos}. \Cref{thm:depth-3-subexp} shows that one additional layer of $\MOD_p$ gates makes such circuits much more powerful. 

It is well-known that for distinct primes $p,q$, every symmetric function on $n$ bits has a $\MOD_{pq}$ circuit of size $\exp(O(n^{\eps}))$ and depth $O(1/\eps)$.\footnote{The authors don't know the origin of this construction. It follows from the fact that every function on $k$ bits has a depth-$2$ $\MOD_{pq}$ circuit of size $2^{O(k)}$, and that symmetric functions can be easily ``decomposed'' into smaller functions (as in~\cite{AllenderKoucky}).} Our result shows the depth can always be made $3$, at the cost of increasing the modulus to a large enough constant. Hansen~\cite{Hansen-CCC06}, building on Bhatnagar, Gopalan, and Lipton~\cite{BhatnagarGL06}, shows that for $m$ which is the product of $r$ primes, and sufficiently small $\ell$ (smaller than each of the prime factors of $m$), the $\MOD_{\ell}$ function can be represented by a polynomial over $\Z_m$ of degree $O(n^{1/r})$.
As a corollary of Hansen's work, Gopalan observed~\cite{Gopalan06} that for every $\eps > 0$ there is an $m$ such that the $\MOD_2$ function has depth-$3$ $\MOD_m$ circuits of size $2^{n^{\eps}}$.
This naturally suggests the question of whether every symmetric function admits such a circuit, which is answered by our \Cref{thm:depth-3-subexp}. 

Allowing larger depths, we can obtain $\MOD_m$ circuits with a interesting size-depth tradeoff.

\begin{theorem}\label{thm:cc0-symmetric}
Let $d \ge 3$ be an integer, and let $m$ be a product of $r \ge 2$ distinct primes.
Then every symmetric function on $n$ bits can be computed by depth-$d$ $\text{MOD}_m$ circuits of size $\exp(\tilde{O}(n^{1/(r+d-3)}))$.
\end{theorem}

To contrast, recall that the lower bounds for $\AC^0$ are $\exp(\Omega(n^{1/(d-1)}))$ size for depth $d$~\cite{Hastad86}, and the lower bounds for $\AC^0[p^k]$ (with AND, OR, and $\MOD_{p^k}$ gates) are $\exp(\Omega(n^{1/(2d)}))$~\cite{Razborov87,Smolensky87} for prime power $p^k$ (where the constant factor depends on $p^k$).  
Thus for constant moduli $m$ with enough prime factors, one can beat both lower bounds with $\MOD_m$ gates. 

For large enough depth $d$, we can achieve even smaller circuits with a size bound of the form $\exp(n^{O(1)/(r\cdot d)})$, multiplying $r$ and $d$ in the denominator, instead of adding them. 

\begin{theorem}\label{thm:cc0-symmetric2} There is a universal constant $c \geq 1$ such that, for all sufficiently large 
depths $d$, and all composite $m$ with $r$ prime factors, every symmetric function can be computed by a $\MOD_m$ gate circuit of depth $d$ and size $\exp(O(n^{c/((d-c)(r-1))}))$.
\end{theorem}

We remark that the constant $c$ in the above construction is not terribly small.\footnote{Our $c$ is at least $6$; this matters if one cares about very small $d$ and $r$.}
In concurrent (very recently released) work,~\cite{idziak2021complexity} give a circuit construction with a similar tradeoff (but better constants) for the special case of the AND function, building on the polynomials of~\cite{BarringtonBR94}.

\paragraph{Even Smaller Circuits in ACC0.} Allowing AND and OR gates in our circuit, the size of our circuit constructions can be further improved. Say that a product $m$ of primes $q_1,\ldots,q_r$ is {\bf good} if every prime factor of $\phi(m)$ divides $m$. We note that the primorial $m = p_r\#$, the product of the first $r$ primes, is good.\footnote{Indeed, for all $i=1,\ldots,r$, the prime factors of $q_i-1$ are contained in $\{q_1,\ldots,q_{i-1}\}$, all of which divide $m=p_r\#$.}

\begin{theorem}\label{thm:acc-symmetric}
Let $m$ be a good product of $r$ primes. 
For every symmetric function $f$ on $n$ inputs and every depth $d \ge 4$ congruent to $1$ modulo $3$, there exists an $\mathsf{AC}^0[m]$ circuit of depth $d$ and size $\exp(\tilde{O}(n^{3/((r+3)(d-1)-3)}))$ computing $f$.
\end{theorem}

In the proof of Theorem~\ref{thm:acc-symmetric}, we make use of several tools from the recent $\AC^0[2]$ circuits of~\cite{OliveiraSS19} (circuits for elementary symmetric polynomials and circuits for the coin problem), along with known results on computing elementary symmetric polynomials modulo a prime. 

Applying standard tricks (seen in~\cite{HP10,WilliamsSTOC14,OliveiraSS19}), Theorem~\ref{thm:acc-symmetric} extends to linear threshold functions.

\begin{corollary}\label{cor:LTF}
Let $m$ be a good product of $r$ primes. 
For every linear threshold function $f$ on $n$ inputs and every depth $d \ge 4$ congruent to $1$ modulo $3$, there exists an $\mathsf{AC}^0[m]$ circuit of depth $d+2$ and size $\exp(\tilde{O}(n^{3/((r+3)(d-1)-3)}))$ computing $f$.
\end{corollary}

This follows directly from the fact that every linear threshold function can be written as an OR of $\poly(n)$ ANDs of $\poly(n)$ symmetric functions on $n$-bit inputs~\cite{HP10}. In general, Theorem~\ref{thm:acc-symmetric} implies that $\TC^0$ circuits (composed of MAJORITY and NOT gates) with small fan-in also have a nontrivial simulation.

\begin{corollary} Every $\TC^0$ circuit of depth $e$ in which every gate has fan-in at most $s$ has an equivalent $\MOD_m$ circuit of depth $d\cdot e$ and size at most $\exp(\tilde{O}(s^{3/((r+3)(d-1)-3)}))$, where $m$ is a good product of $r$ primes.
\end{corollary}

The corollary follows from direct substitution of each MAJORITY gate with depth-$d$ circuits from our Theorem~\ref{thm:acc-symmetric}. Note that such depth-$e$ $\TC^0$ circuits have at most $O(s^{e-1})$ gates.

\paragraph{Can't you do any better?} Theorem~\ref{thm:acc-symmetric} shows that for certain $m$ which are products of $r$ primes, one can compute arbitrary symmetric functions in depth $d$ and size $\exp(n^{\frac{c}{rd}})$ where $c > 0$ is a constant. We give evidence that it may be difficult to improve asymptotically on the dependence of $r$ and $d$ in the exponent of $n$, based on a natural hypothesis 
regarding $\TC^0$ circuits, which are constant-depth circuits composed of MAJORITY and NOT gates.
(Of course it is difficult to prove anything unconditional here, because as far as we know, polynomial-size depth-$3$ $\MOD_6$ circuits could compute every $\mathsf{EXP}$ function. Thus we settle for conditional hardness.)

Recall that a $\SYM \circ \AND$ circuit is a depth-two circuit where the output is a symmetric function and the bottom layer computes ANDs of input variables and negations. 
The hypothesis is that subexponential-size $\SYM \circ \AND$ circuits cannot compute $\TC^0$ circuits in which each gate has linear fan-in. 

\begin{hypothesis}[$\SYM \circ \AND$ Hypothesis] \label{hyp:sym-and}
There are constants $c, k > 1$ such that for sufficiently large $n$, there is a function $f : \{0,1\}^n \rightarrow \{0,1\}$ computable by $\TC^0$ circuits of depth $c$ with at most $\tilde{O}(n)$ gates where each gate has fan-in $\tilde{O}(n)$, such that $f$ does not have an $\exp(O(n^{1/k}))$ size $\SYM \circ \AND$ circuit.
\end{hypothesis}

A well-known result in circuit complexity is that every $\ACC^0$ circuit of size $s$ can be simulated by a $\SYM \circ \AND$ circuit of size $s^{\poly(\log s)}$~\cite{BeigelTarui,ChenP19}. Therefore, the $\SYM \circ \AND$ Hypothesis is a strengthening of the longstanding hypothesis that $\TC^0 \not\subset \ACC^0$: the $\SYM \circ \AND$ Hypothesis implies exponential lower bounds for simulating $\TC^0$ circuits with $\ACC^0$ circuits. Indeed, the hypothesis implies that our $\ACC^0$ circuits for symmetric functions are nearly size-optimal in their dependence on depth and modulus.

\begin{theorem}[Near-Optimality Modulo a Conjecture]
\label{thm:optimal} Assuming the $\SYM \circ \AND$ Hypothesis, there is a fixed $\alpha > 0$ such that for \emph{every} $m$ and $d$, every depth-$d$ $\ACC^0[m]$ circuit computing the MAJORITY function on $n$ inputs requires size at least $\exp(n^{\frac{\alpha}{rd}})$ for sufficiently large $n$, where $r$ is the number of distinct prime factors of $m$.
\end{theorem}

The proof of \Cref{thm:optimal} is in Appendix~\ref{sec:optimal}.
Therefore, we view size bounds of the form $\exp(n^{1/\Theta(rd)})$ (as seen in our results) as a natural barrier to better upper bounds on $\MOD_m$ circuits: any function with significantly smaller $\MOD_m$ circuit complexity (as a function of $n$, $r$, and $d$) would also yield a highly non-trivial $SYM \circ \AND$ circuit simulation of $\TC^0$. In order to achieve significantly smaller circuits as a function of $n$, $d$, and $r$, one has to at least refute the hypothesis. Of course, even assuming Hypothesis~\ref{hyp:sym-and}, our circuits can probably be improved by constant factors in the exponents. 

\section{Preliminaries}

We assume basic familiarity with computational complexity~\cite{Arora-Barak09} and Boolean circuit complexity~\cite{Jukna12}, although we have tried to keep the paper relatively self-contained.

\paragraph{Notation.} For a binary vector $\mathbf{x}$, we use $|\mathbf{x}|_1$ to denote the $\ell_1$-norm, i.e., the number of ones in $\mathbf{x}$.

Besides $\AC^0$, $\ACC^0$, $\CC^0$, and $\TC^0$, we also use the following additional notation for various circuit types, all of which is standard. A circuit of type $\SYM$ is simply a symmetric Boolean function. An EMAJ function outputs $1$ on an input $(x_1,\ldots,x_n) \in \{0,1\}$ if and only if $\sum_i x_i = T$ for a fixed target $T$. A circuit of type $G_1 \circ \cdots \cdot G_d$ notation is a circuit where the output gate has type $G_1$, the next layer of gates all have type $G_2$, and so on, and the bottom layer (nearest the inputs) only contains gates of type $G_d$.

The following basic fact is useful to keep in mind.

\begin{proposition} For all positive $m, n \in \Z$, any $\MOD_m$ gate of fan-in $t$ can be simulated by a $\MOD_{mn}$ of fan-in $nt$.
\end{proposition}

\begin{proof} For any positive integer $t$, $m \mid t$ if and only if $nm \mid nt$. So $\MOD_m(x_1,\ldots,x_t) = \MOD_{mn}(n \cdot x_1,\ldots, n \cdot x_t)$. \end{proof}

\paragraph{Tools.} We make use of several known results. First, we note that AND circuits of small fan-in have efficient depth-two $\MOD_m$ circuits. A version was first used in~\cite{Barrington-Immerman-Straubing90} in the context of MOD circuits, and more recently a strengthening was used to reduce the size-depth tradeoff for simulating $\ACC^0$ circuits with $\text{SYM} \circ \text{AND}$ circuits~\cite{ChenP19}. (Chen and Papakonstantinou~\cite{ChenP19} call this ``linearization''.)

\begin{proposition}[\cite{Barrington-Immerman-Straubing90,ChenP19}] 
\label{prop:AND-representation}
Let $a,b \geq 2$ be fixed integers with $gcd(a,b) = 1$. Every AND of $k$ $\MOD_b$ gates can be represented by an $\MOD_a \circ \MOD_b$ circuit of $O(b^{k})$ gates. Furthermore, on all $k$-bit inputs, the sum of the inputs to the output gate of the circuit is always $0 \pmod a$ or $1 \pmod a$.
\end{proposition}

Our next tool is an old number-theoretic theorem on elementary symmetric polynomials modulo $p$, masterfully applied by Beigel, Barrington, and Rudich~\cite{BarringtonBR94} in their non-trivial degree  polynomials for the OR functions over composite moduli.

\begin{theorem}[Lucas' Theorem~\cite{Lucas78}]
\label{thm:lucas}
For all primes $p$ and natural numbers $n$, 
\[\binom{n}{p^i}\bmod p\] is the $i$-th digit in the $p$-ary representation of $n$.
\end{theorem}

Lucas' theorem has the following direct consequence for polynomial representations of Boolean functions. 

\begin{lemma}[\cite{BarringtonBR94}] \label{lemming1}
Let $p$ be a prime, let $n$ be a natural number, and let $e_i(\mathbf{x})$ denote the
$i$-th
elementary symmetric polynomial on $n$ variables.
For a binary vector $\mathbf{x}$, let \[\displaystyle \sum y_i \cdot p^i = |\mathbf{x}|_1\] be the $p$-ary expansion of $\displaystyle |\mathbf{x}|_1$.
Then for every $i$, $e_{p^i}(\mathbf{x}) \equiv y_i {\bmod p}$.
\end{lemma}

In order to apply the elementary symmetric polynomials, our construction also involves arithmetic circuits over prime fields. These circuits will be translated into Boolean circuits with $\MOD_m$ gates.

\begin{lemma}[\cite{ChenOST16,OliveiraSS19}] \label{lemming2}
Let $p$ be a prime, let $n, i \in \mathbb{N}$, and let $d \ge 2$ be even.
There is an arithmetic circuit over $\mathbb{F}_p$ of depth $d$ and size $n^{O(i^{2/d})}$ computing the 
$i$-th
elementary symmetric polynomial (over $\mathbb{F}_p$) on $n$ inputs, where the output gate is a $\times$ gate.
\end{lemma}

We also use $\AC^0$ circuits for the coin problem. These were also used by~\cite{OliveiraSS19} in their improved $\AC^0[2]$ circuits for symmetric functions. 

In the following, we let $i, j \in \{0,1,\ldots,n\}$, and let $D_{i,j}$ be any partial function satisfying the properties:
\[\text{$D_{i,j}(\mathbf{x}) = 1$ if $|\mathbf{x}|_1 = i$, and}\]
\[\text{$D_{i,j}(\mathbf{x}) = 0$ if $|\mathbf{x}|_1 = j$}.\] 

\begin{lemma}[\cite{ODonnellW07,Amano09,OliveiraSS19}] \label{lemming3}
Let $d \ge 2$ and $n$ be natural numbers, and let $i \ne j$.
Then there is an $\mathsf{AC}^0$ circuit of depth $d$ and size $\displaystyle \exp({ O ( d \left( n/|i-j| \right)^{1/(d-1)} ) })$ computing $D_{i,j}$ on $n$ inputs, where the output gate is an $\text{AND}$.
\end{lemma}

Intuitively, \Cref{lemming3} will be useful when $|i-j|$ is ``large''. 

\section{CC0 Circuits for Symmetric Functions}

We begin by giving efficient depth-3 $\CC^0$ circuits for symmetric functions.

\begin{reminder}{\Cref{thm:depth-3-subexp}}
For every $\eps > 0$, there is a modulus $m \le (1/\eps)^{2/\eps}$ such that every symmetric function on $n$ bits can be computed by depth-$3$ $\MOD_m$ circuits of $\exp(O(n^{\eps}))$ size. In fact, the circuits have the form $\MOD_{p_1} \circ \MOD_{p_2 \cdots p_r} \circ \MOD_{p_1}$, where $p_1,\ldots,p_r$ are distinct primes.
\end{reminder}

After that, we will generalize the result to a size-depth tradeoff in the next subsection. 
That tradeoff will be further improved in \Cref{sec:AC0-tradeoff} when we allow the use of $\text{AND}$ and $\text{OR}$ gates.

As a warm-up, we first consider the special case where $\eps > 1/3$ and $m=30$.

\begin{theorem}\label{thm:special-case-30} Every symmetric Boolean function on $n$ variables has a depth-three circuit of the form $\MOD_5 \circ \MOD_6 \circ \MOD_5$, of size $\exp(O(n^{1/3} \log n))$. Furthermore, the output gate is a linear sum which always evaluates to either $0$ or $1$ modulo $5$. 
\end{theorem}

Note that the upper bound of \Cref{thm:special-case-30} already beats the well-known lower bounds for depth-3 $\AC^0$~\cite{Hastad86}.
The remainder of this section is devoted to the proof.
A key component is a low-degree multivariate polynomial over $\Z_6$ that vanishes on a Boolean vector if and only if the sum of the ones in the vector equals a particular value.

\begin{theorem} \label{thm:exactmaj-cc}
For every $n \in \mathbb{N}$ and every $T \in \{0,1,\ldots,n\}$, there is a polynomial $P_T(x_1,\ldots,x_n)$ of degree at most $3\sqrt{n}$ such that for all $a \in \{0,1\}^n$, $P_T(a) = 0 \bmod 6$ if and only if $\displaystyle \sum_i a_i = T$.
\end{theorem}

\begin{proof} We want a polynomial $p$ on $n$ variables such that for all $y_1,\ldots,y_n \in \{0,1\}$ and $T \in \{0,1,\ldots,n\}$,
\[p(y_1,\ldots,y_n) \equiv 0 \bmod 6 \iff \sum_i y_i = T.\]

For the elementary symmetric polynomial $e_J(y_1,\ldots,y_n)$ of degree $J$, and for all $a_1,\ldots,a_n \in \{0,1\}$, 
\[e_J(a_1,\ldots,a_n) = \binom{(\sum_i a_i)}{J}.\]
Thus by Lucas' Theorem~(\Cref{thm:lucas}), $e_{p^i}(a_1,\ldots,a_n) \bmod p$ equals the $i$-th digit in the $p$-ary representation of $\sum_i a_i$.

Let $s$ and $t$ be integers so that $2\sqrt{n} \geq 2^s > \sqrt{n}$ and $3\sqrt{n} \geq 3^t > \sqrt{n}$.

Suppose when we write $T \in \{0,1,\ldots,n\}$ in binary notation, the $s$ low order bits are $b_{s-1}, \ldots, b_0$. Furthermore, when we write $T$ in ternary notation, the $t$ low order trits are $c_{t-1},\ldots, c_0$.

Define the polynomials 
\[p_2(y_1,\ldots,y_n) := 1 - \prod_{j=0}^{s-1} \left(1-\left(b_j - e_{2^j}(y)\right)\right) \bmod 2\] and
\[p_3(y_1,\ldots,y_n) := 1 -  \prod_{j=0}^{t-1} (1-(c_j - e_{3^j}(y))^2) \bmod 3.\]

Note the degrees of $p_2$ and $p_3$ are $O(\sqrt{n})$. In particular, $\displaystyle \deg(p_2) = \sum_{j=0}^{s-1} 2^j = 2^s - 1$ and $\displaystyle \deg(p_3) = \sum_{j=0}^{s-1} (2\cdot 3^j) = 2(3^s-1)/2 = 3^s - 1$.

We observe a few properties of the polynomials $p_2$ and $p_3$:

\begin{proposition} For all $a \in \{0,1\}^n$, $p_2(a)\equiv 0\bmod 2$ if and only if the binary representation of $\sum_i a_i$ equals $b_{s-1} \cdots b_0$ in the last $s$ bits. Analogously, $p_3(a) \in \{0,1\} \bmod 3$, and $p_3(a) \equiv 0 \bmod 3$ if and only if the ternary representation of $\sum_i a_i$ equals $c_{t-1}\cdots c_0$ in the last $t$ trits.
\end{proposition}

\begin{proof} We prove the proposition for $p_3$; the case of $p_2$ is analogous. Let $a \in \{0,1\}^n$. Each difference $(c_j - e_{3^j}(a))^2$ is either $0$ or $1$ modulo $3$, and it is $0$ if and only if $c_j = e_{3^j}(a)$. Thus the product 
$\prod_{j=0}^{t-1} (1-(c_j - e_{3^j}(a))^2)$ equals $1$ if and only if $c_j = e_{3^j}(a)$ for all $j=0,\ldots,t-1$, hence $p_3(a) \equiv 0 \bmod 3$ if and only if $c_j \equiv e_{3^j}(a) \bmod 3$ for all $j=0,\ldots,t-1$. Recalling that $(e_{3^j}(a) \bmod 3)$ equals the $j$-th trit of $\sum_i a_i$, the result follows.
\end{proof}

We note that in general, working modulo a prime $q$, we may construct a polynomial with degree $(q^{t}-1)$ of the form
\begin{equation}
\label{eq:general-poly}
p_q(y_1,\ldots,y_n) = 1 -  \prod_{j=0}^{t-1} (1-(c_j - e_{q^j}(y))^{q-1}).
\end{equation}

By the above proposition, it follows that for all $a \in \{0,1\}^n$,
\[p_2(a) \equiv 0 \bmod 2 \iff \sum_i a_i \equiv T \bmod 2^s\]
and 
\[p_3(y) \equiv 0 \bmod 3 \iff \sum_i a_i \equiv T \bmod 3^t.\]

Since $\sum_i a_i$ and $T$ are both in $\{0,\ldots,n\}$ and $2^s \cdot 3^t > n$, by the Chinese Remainder Theorem we have
\begin{align*}
\sum_i a_i = T &\iff (\sum_i a_i \equiv T \bmod 2^s) \wedge
(\sum_i a_i \equiv T \bmod 3^t)\\
&\iff (p_2(y) \equiv 0 \bmod 2)\wedge(p_3(y) \equiv 0 \bmod 3)\iff 3p_2(y) + 2p_3(y) \equiv 0 \bmod 6.
\end{align*}
Thus $3p_2(y) + 2p_3(y)$ is a polynomial of degree $O(\sqrt{n})$ which equals $0 \bmod 6$ if and only if $\displaystyle \sum_i y_i = T$.
This completes the proof of \Cref{thm:exactmaj-cc}.
\end{proof}

We now proceed with the proof of \Cref{thm:special-case-30}.

\begin{proof}
Let $f$ be a symmetric function and let $g : \{0,1,\ldots,n\} \rightarrow \{0,1\}$ be its companion function. That is, for every $\mathbf{x}$, $f(\mathbf{x}) = g(|\mathbf{x}|_1)$.

The output gate will be a $\MOD_5$ gate that \begin{itemize}
    \item (a) sums over possible choices of $T \in \{0,1,\ldots,n\}$ such that $g(T) = 1$ and 
    \item (b) sums over all ways to partition $T$ into a sum of $t= \lceil n^{1/3} \rceil$ parts $T_1,\ldots,T_t \in \{0,1,\ldots,T\}$.
\end{itemize}
 There are $2^{O(n^{1/3} \log n)}$ choices over (a) and (b). We associate each part $T_i$ with a disjoint set $S_i$ of at most $\lceil n^{2/3} \rceil$ variables from the input. For each of the choices from (a) and (b), we wish to verify that, for all $i=1,\ldots,t$, the sum of all variables in $S_i$ equals $T_i$. Note that there is \emph{at most} one choice from (a) and from (b) that could possibly be consistent with the given input, so we can use a \emph{modulo-$5$ sum} (not just a $\MOD_5$ gate) to sum over these choices. This modulo-$5$ sum will always be either $0$ or $1$ modulo $5$.

By our construction of EMAJ polynomials, each sum over the set $S_i$ of $n^{2/3}$ variables can be checked with a $\MOD_6$ gate of $2^{O(n^{1/3})}$ fan-in, where each input to the $\MOD_6$ gate is the output of an AND of fan-in $O(n^{1/3})$. Putting these $\MOD_6 \circ AND$ circuits below each wire of the modulo-5 sum, at this point, we have a modulo-5 sum of $2^{O(n^{1/3} \log n)}$ ANDs of fan-in $O(n^{1/3})$ of $\MOD_6$ of fan-in $2^{O(n^{1/3})}$ of ANDs of fan-in $O(n^{1/3})$. 

To eliminate the AND gates, we apply Proposition~\ref{prop:AND-representation}, yielding that an AND of $f$ MOD$q$ gates can be represented by a modulo-$p$ sum of $O(q^f)$ MOD$q$ gates, as long as $\gcd(p,q)=1$. In particular, for the ``middle'' ANDs we set $p=5$ and $q=6$, and for the ``bottom'' ANDs we set $p=6$ and $q=5$. We obtain a modulo-5 sum of $2^{O(n^{1/3} \log n)}$ $\MOD_6$ of fan-in $2^{O(n^{1/3})}$ of $\MOD_5$ of fan-in $O(n^{1/3})$.
\end{proof}

The above construction has several interesting corollaries; here is one.

\begin{corollary}
Every circuit of the form $\text{MOD}_5 \circ \text{SYM}$ of size $2^{O(n^{1/3} \log n)}$ can be simulated by a depth-three $\text{MOD}_5 \circ \text{MOD}_6 \circ \text{MOD}_5$ circuit of size $2^{O(n^{1/3}\log n)}$.
\end{corollary}

\begin{proof} We simply replace each SYM gate (which takes $n$ inputs) in the $\text{MOD}_5 \circ \text{SYM}$ circuit with a modulo-5 sum of $\text{MOD}_6 \circ \text{MOD}_5$ as in the previous theorem. 
\end{proof}

We are now ready to generalize to \Cref{thm:depth-3-subexp}.

\begin{reminder}{\Cref{thm:depth-3-subexp}}
For every $\eps > 0$, there is a modulus $m \le (1/\eps)^{2/\eps}$ such that every symmetric function on $n$ bits can be computed by depth-$3$ $\MOD_m$ circuits of $\exp(O(n^{\eps}))$ size. In fact, the circuits have the form $\MOD_{p_1} \circ \MOD_{p_2 \cdots p_r} \circ \MOD_{p_1}$, where $p_1,\ldots,p_r$ are distinct primes.
\end{reminder}

\begin{proof}
Let $\eps > 0$, and let $f$ be a symmetric function. Take $k := \lfloor 1+1/\eps \rfloor$, let $m$ be the product of the first $k$ primes, and let $m' = m/2$.

We use a similar construction as in \Cref{thm:special-case-30} to get a $\text{MOD}_2 \circ \text{MOD}_{m'} \circ \text{MOD}_2$ circuit for $f$.

The differences are that we partition the target $T \in \{0,1,\ldots,n\}$ into a sum of $\lfloor n^{1/k} \rfloor$ parts where each part is over $v := n^{1-1/k}$ variables, and by using $k-1$ primes instead of two, we can obtain a polynomial for $\text{EMAJ}$ on $v$ variables of degree $O(v^{1/(k-1)}) \leq O(n^{1/k})$ in an analogous way. 

More precisely, let $T' \in \{0,1,\ldots,v\}$ be a target value. For the first $k-1$ odd primes $q_1,\ldots,q_{k-1}$, we take $k-1$ polynomials $p_{q_1}(x),\ldots,p_{q_{k-1}}(x)$ as defined in \eqref{eq:general-poly} such that each $p_{q_i}(x)$ has degree $(q_i)^{t_i}-1$, where the $t_i$ are chosen such that for all $i \in [k]$, 
\begin{itemize}
    \item $(q_i)^{t_i} = \Theta(v^{1/(k-1)})$,
    \item $T' \leq v < \prod_i (q_i)^{t_i}$, and
    \item for all $a \in \{0,1\}^v$ we have \[p_{q_i}(a) = 0 \bmod q_i \iff \sum_i a_i \equiv T' \bmod (q_i)^{t_i}.\] 
\end{itemize}

By the Chinese Remainder Theorem, and similar reasoning as in \Cref{thm:special-case-30}, there are fixed coefficients $M_i \in [m']$ such that
\[ \sum_{i=1}^{k-1} M_i \cdot p_{q_i}(a) \equiv 0 \bmod m' \iff \bigwedge_{i=1}^{k-1} \left[p_{q_i}(a) \equiv 0 \bmod q_i\right] \iff \sum_i a_i = T'.\] Thus we have a polynomial of degree $O(v^{1/(k-1)})$ that vanishes modulo $m'$ precisely when the sum of $v$ variables equals the target $T'$. Naturally we might write this polynomial as a $\MOD_{m'} \circ \AND$ circuit of $\exp(\tilde{O}(v^{1/(k-1)}))$ size; by replacing each $\AND$ with a modulo-$m'$ sum of $\MOD_2$ gates (\Cref{prop:AND-representation}), we can express it as a $\MOD_{m'} \circ \MOD_2$ circuit. 
Our final circuit has the form $\MOD_2 \circ \MOD_{m'} \circ \MOD_2$ and size $\exp(\tilde{O}(n^{1/k})) \leq \exp(O(n^{\eps}))$.
\end{proof}

\subsection{Size-Depth Tradeoff with CC0}

Allowing depth $d$ circuits for $d > 3$, the size of the above construction can be improved as a function of the number of distinct primes $r$ in the modulus. Here we only briefly describe the construction, as the size bound will be improved significantly (as a function of $d$ and $r$) in the following section.

\begin{reminder}{\Cref{thm:cc0-symmetric}}
Let $d \ge 3$ be an integer, and let $m$ be a product of $r \ge 2$ distinct primes.
Then every symmetric function on $n$ bits can be computed by depth-$d$ $\text{MOD}_m$ circuits of size $\exp({\tilde{O}(n^{1/(r+d-3)})})$.
\end{reminder}

\begin{proof} Let $p$ and $q$ be the smallest prime factors of $m$. We prove by induction on $d$ that there are circuits of size $\exp(O(n^{1/(r+d-3)}\log n))$, and we prove additionally that the output gate is a $\text{MOD}_p$ gate when $d$ is odd and a $\text{MOD}_q$ gate when $d$ is even.

For $d=3$, we use the construction of \Cref{thm:depth-3-subexp} to obtain a $\text{MOD}_p \circ \text{MOD}_{m/p} \circ \text{MOD}_p$ circuit of depth $3$ and size $\exp(O(n^{1/r}\log n))$.

For the inductive step, we proceed similarly to the proof of \Cref{thm:depth-3-subexp}, except that we use a $\text{MOD}_p$ or $\text{MOD}_q$ gate as the output gate (depending on the parity of $d$). We partition the target $T$ into a sum of $t = \lceil n^{1/(r+d-3)} \rceil$ parts, where each part contains at most $\lceil n^{(r+d-4)/(r+d-3)} \rceil$ variables, and our circuit sums over all $\exp(O(t \log n))$ choices for the number of true variables in each part. Since $\text{EMAJ}$ is a symmetric function, we can inductively compute each $\text{EMAJ}$ on $\lceil n^{(r+d-4)/(r+d-3)} \rceil$ variables with a circuit of depth $d-1$, as guaranteed by the inductive hypothesis. These circuits have size $\exp(O((n^{(r+d-4)/(r+d-3)})^{1/(r+d-4)} \log n)) = \exp(O( n^{1/(r+d-3)} \log n))$, and their output gates have fan-in $\exp(O( n^{1/(r+d-3)} \log n))$. 
WLOG assume $d$ is odd. Then the depth-$(d-1)$ circuits for $\text{EMAJ}$ described above have the form
\[\MOD_q \circ \cdots \circ \MOD_p \circ \MOD_{m/p} \circ \MOD_p,\] and our entire circuit has the form  \[\MOD_p \circ \AND \circ \MOD_q \circ \cdots \circ \MOD_p \circ \MOD_{m/p} \circ \MOD_p,\] where the $\AND$s have fan-in $t$. As before, each $\AND \circ \MOD_q$ can be replaced by a modulo-$p$ sum of $\MOD_q$ gates using Proposition~\ref{prop:AND-representation}, which only increases the circuit size by a factor of $2^{O(t)}$.

Our final circuit has depth $d$ and size $\exp(O(n^{1/(r+d-3)} \log n))$.
\end{proof}

\paragraph{A Better Dependence on Depth and Modulus.} We can give a $\CC^0$ circuit construction with a better asymptotic tradeoff (in the double-exponent). We will keep the description of this construction brief and to the point, as its size will be further improved (replaced by better constants) in the next section, using OR and AND gates.

\begin{reminder}{Theorem~\ref{thm:cc0-symmetric2}} There is a universal constant $c \geq 1$ such that, for all sufficiently large 
depths $d$, and $m$ which is the product of the first $r$ prime factors, every symmetric function can be computed by a $\MOD_m$ gate circuit of depth $d$ and size $\exp(O(n^{c/((d-c)(r-1))}))$.
\end{reminder}

\begin{proof} First, we recall that every symmetric function $f$ on $n$ variables can be expressed as a MAJORITY of $O(n)$ MAJORITY gates over the $n$ variables~(see for example~\cite{BeameBL92} for a reference). Thus it suffices to give a circuit for MAJORITY.

Allender and Koucky~\cite[Theorem~3.8]{AllenderKoucky} give a downward self-reduction for the MAJORITY function: they prove that there is a universal constant $a \geq 1$ such that for every $k \geq 1$, the MAJORITY function on $n$ bits can be computed by a $\TC^0$ circuit of depth at most $ak$ where each MAJORITY gate has fan-in at most $O(n^{1/k})$. Applying these circuits to the depth-two $\TC^0$ circuits described in the previous paragraph, we obtain an analogous circuit of depth $2ak$ for any given symmetric function $f$. 

Replace each MAJORITY gate of fan-in at most $O(n^{1/k})$ with a depth-3 $\MOD_m$ circuit of size at most \[\exp(\tilde{O}(n^{1/(k(r-1))})),\] as provided by Theorem~\ref{thm:cc0-symmetric}. (Note that each NOT gate can always be replaced by a single $\MOD_m$ gate, if we do not want to allow NOT gates in our $\CC^0$ circuit.) This results in a circuit of depth $6ak$ and size \[\exp(\tilde{O}(n^{1/(k(r-1))})) \leq \exp(\tilde{O}(n^{1/(k(r-1))})).\] Thus for depths $d = 6ak$ where $k$ is a positive integer, the size bound is at most $\exp(n^{6a/(d(r-1))})$. 
For depths $d$ that are not divisible by $6a$, we can simply use the construction for $d' = 6ak$ where $d' < d < 6a(k+1)$, which has size at most $\exp(n^{6a/(d'(r-1))}) < \exp(n^{6a/((d-6a)(r-1))})$. 
\end{proof}

The above construction is not useful for $d < 6$ and small $r$, which are of interest. In the next section, we will show that much better constants are obtainable in the $\ACC^0$ setting. 

\section{Size-Depth Tradeoff With ACC0}\label{sec:AC0-tradeoff}

We now turn to showing how adding AND and OR gates can help improve the circuits even further. We begin with a result using the concrete modulus $42$.

\begin{theorem}\label{tradeoff-AC0[42]}
For every symmetric function $f$ on $n$ inputs and every depth $d$ with $d \equiv 2 \bmod 6$, there exists an $\mathsf{AC}^0[42]$ circuit of depth $d$ and size $\displaystyle \exp(\tilde{O}(n^{\frac{6}{13(d-2)}}))$ computing $f$.
\end{theorem}

Observe that, for sufficiently large $d$, the circuit size of \Cref{tradeoff-AC0[42]} already drops below Smolensky's $\AC^0[p^k]$ depth-$d$ lower bound of $\exp(\Omega(n^{1/(2d)}))$ size~\cite{Smolensky87} for computing $\MOD_q$ when $gcd(p,q)=1$.

We build on the results of Oliveira et al.~\cite{OliveiraSS19} for computing symmetric functions in $\mathsf{AC}^0[2]$.
At a high level, we note that every symmetric function can be written as an OR of AND of (partial) functions of the form $D_{i,j}$, where 
\[\text{$D_{i,j}(\mathbf{x}) = 1$ if $|\mathbf{x}|_1 = i$, and}\]
\[\text{$D_{i,j}(\mathbf{x}) = 0$ if $|\mathbf{x}|_1 = j$},\] recalling that $|\mathbf{x}|_1$ is the number of $1$'s in $\mathbf{x}$. Note that $D_{i,j}$ could have \emph{arbitrary} behavior on any other Boolean inputs.

When $|i-j|$ is large, the function $D_{i,j}$ can be simulated by the standard Coin Problem, for which there are known $\mathsf{AC}^0$ circuits (see the Preliminaries). When $|i-j|$ is small, we give a new construction of $\mathsf{AC}^0[42]$ circuits for $D_{i,j}$.

We will utilize arithmetic circuits for elementary symmetric polynomials. To that end, the following lemma shows how to generically translate low-depth arithmetic circuits over $\F_p$ into $\AC^0[p(p-1)]$ circuits, in a way that only increases the circuit depth by a $3/2$ multiplicative factor. (Getting some constant factor increase is not too difficult; \cite{AgrawalAD00} first showed a correspondence between $\ACC^0$ and arithmetic circuits over finite fields.)

\begin{lemma}\label{lemming4}
Let $p$ be prime, and let $C$ be an arithmetic circuit over $\mathbb{F}_p$ of size $s$ and depth $2d$ (with alternating layers of $+$ and $\times$ gates) on $n$ inputs, such that for every $\mathbf{x} \in \{0,1\}^n$, $C(\mathbf{x}) \in \{0,1\}$.
Then $C$ is equivalent to an $\mathsf{AC}^0[p(p-1)]$ circuit $C'$ of size $O(s\cdot p)$ and depth $3d$. 
\end{lemma}
\begin{proof}
We represent an element $x$ of $\mathbb{F}_p$ in unary, by  $p$ indicator bits \[b_0(x), b_1(x), \ldots, b_{p-1}(x),\] where $b_0(x) = 0$ iff $x = 0$, and for $i \ne 0$, we let $b_i(x) = 1$ iff $x = i$.
(We treat the $0$-th indicator bit as a special case to make later constructions easier.)
We now obtain $C'$ by replacing each gate in $C$ with a small $\mathsf{AC}^0[p(p-1)]$ gadget circuit.

For each addition gate of $C$ computing \[\displaystyle x = \sum_{j=1}^k x_j,\] we replace that gate with $p$ parallel $\text{MOD}_p$ gates, so that
\[\displaystyle b_i(x) = 1 \iff (p-i) + \sum_{j=1}^k \sum_{i'=1}^{p-1} i' \cdot  b_{i'}(x_j) \equiv 0 \bmod p.\]
(As a special case, we output the negation of the right hand side in the case of $b_0(x)$.)
To see why this works, we observe that the inner sum computes $x_j$, and so the outer sum computes $x$.
Now $x+(p-i) \equiv 0 \bmod p$ precisely when $x = i$.

Take $g$ to be a generator of the multiplicative group $\mathbb{F}_p^*$ of $\mathbb{F}_p$, and let  $\log_g(n)$ denote the discrete logarithm base $g$ in $\mathbb{F}_p$  (i.e., $g^{\log_g(n)} = n \bmod p$).
For each multiplication gate of $C$ computing \[\displaystyle x = \prod_{j=1}^k x_j,\] we replace that gate with an $\text{AND}$ gate placed in parallel with $p-1$ $\text{AND} \circ \text{MOD}_{p-1}$ circuits, implementing the conditions
\[\displaystyle b_0(x) = \bigwedge_{j=1}^k b_0(x_j),\]
and for $i \ne 0$,
\[b_i(x) = G_i \wedge \bigwedge_{j=1}^k b_0(x_j).\]
where $G_i$ is a $\text{MOD}_{p-1}$ gate such that
\[G_i = 1 \iff (p - \log_g(i)) + \sum_{j=1}^k \sum_{i'=2}^{p-1} b_{i'}(x_j)\cdot  \log_g(i') \equiv 0 \bmod p-1.\]
To see why this works, we observe that the inner sum computes the discrete logarithm of $x_j$ (for the same reason that the inner sum in the addition case computes $x_j$).
Since $x = \prod x_j$, we have (for non-zero $x$) $\log_g x = \sum \log_g x_j$, so the outer sum computes the discrete logarithm of $x$.
Now $(\log_g x) +(p-\log_g i) \equiv 0 \bmod p-1$ precisely when $x = i$.

Finally, we take the output wire of $C'$ to be the negation of the $b_0$ wire from the output gate of $C$.
\end{proof}

We note that as a special case, an arithmetic circuit over $\mathbb{F}_2$ can be viewed directly as an $\mathsf{AC}^0[2]$ circuit (with the same size and depth), since an element of $\mathbb{F}_2$ is simply a bit, addition in $\mathbb{F}_2$ is $\text{MOD}_2$, and multiplication is $\text{AND}$.
Additionally, when $p-1$ is not square-free, we can improve the modulus in the circuit above.

\begin{lemma}\label{lemming5}
Let $p$ be prime, and let $C$ be an arithmetic circuit over $\mathbb{F}_p$ of size $s$ and depth $2d$ (with alternating layers of $+$ and $\times$ gates) on $n$ inputs, such that for every $\mathbf{x} \in \{0,1\}^n$, $C(\mathbf{x}) \in \{0,1\}$.
Then $C$ is equivalent to an $\mathsf{AC}^0[pm]$ circuit of size $O(s^{p-1}p^{(p-1)\log (p-1)})$ and depth $3d$, where $m$ is the product of the distinct prime factors of $p-1$.
\end{lemma}

\begin{proof}
We start with the circuit $C'$ given by Lemma \ref{lemming4}.
We now use Theorem~\ref{thm:lucas} and the Chinese Remainder Theorem to simulate each $\text{MOD}_{p-1}$ gate of fan-in $f$ using an $\text{AND} \circ \text{MOD}_m \circ \text{AND}$, where the three layers of gates have fan-in at most $\log (p-1)$, $f^{p-1}$, and $p-1$, respectively.
The bottom layer of $\text{AND}$ gates have $\text{MOD}_p$ gates as inputs, so can be replaced with a sum (mod $m$) of fan-in $O(p^{p-1})$ using Proposition~\ref{prop:AND-representation}.
This can be absorbed into the layer of $\text{MOD}_m$ gates.
The top layer of $\text{AND}$ gates can similarly be converted into a sum (mod $p$) of fan-in at most $p^{\log (p-1)}$, which can be absorbed into the $\text{MOD}_p$ gates for which they are inputs (since $\text{MOD}_{p-1}$ gates in $C'$ can only be inputs to $\text{MOD}_p$ gates).
\end{proof}

Putting these results together, we obtain the following:
\begin{theorem}\label{tradeoff-MOD}
Let $d$ be a multiple of $6$, let $n$ be a natural number, and let $\alpha \in (0,1]$.
Set \[s := \left\lceil \frac{3 \alpha \log_2 n}{7} \right\rceil,~ t := \left\lceil \frac{2 \alpha \log_3 n}{7} \right\rceil,~ u := \left\lceil \frac{2 \alpha \log_7 n}{7} \right\rceil,\] and $m := 2^s 3^t 7^u$.
Then there is an $\mathsf{AC}^0[42]$ circuit of depth $d+1$ and size $2^{\tilde{O}(n^{6 \alpha / 7d})}$ computing the $\text{MOD}_m$ function on $n$ inputs, where the output gate is an $\text{AND}$ gate.
\end{theorem}

\begin{proof}
Applying Lemma~\ref{lemming2}, we construct: \begin{itemize}
    \item 
arithmetic circuits $C_1, C_2, \ldots, C_{2^s}$ over $\mathbb{F}_2$ of depth $d$, where $C_i$ computes the $i$-th elementary symmetric polynomial modulo $2$,
\item arithmetic circuits $D_1, D_3, \ldots, D_{3^t}$ over $\mathbb{F}_3$ of depth $2d/3$,  where $D_i$ computes the $i$-th elementary symmetric polynomial modulo $3$, and
\item arithmetic circuits $E_1, E_7, \ldots, E_{7^u}$ over $\mathbb{F}_7$ of depth $2d/3$, where $E_i$ computes the $i$-th elementary symmetric polynomial modulo $7$,
\end{itemize}
all of which have size $n^{O(n^{6 \alpha / 7d})}$, given our parameters.

We convert each of the $D_i$ and $E_i$ into $\AC^0[42]$ circuits $D'_i$ and $E'_i$ using Lemma~\ref{lemming4}, and as previously observed, the $C_i$ are already $\AC^0[2]$ circuits.

Finally, from Lemma \ref{lemming1} and the Chinese Remainder Theorem, all of the $C_i(\mathbf{x})$, $D'_i(\mathbf{x})$, and $E'_i(\mathbf{x})$ output $1$ if and only if $|\mathbf{x}|_1 \equiv 0 \bmod m$. Our final circuit for $\MOD_m$ is obtained by taking the $\text{AND}$ of all of these circuits.
\end{proof}

We are now ready to prove Theorem \ref{tradeoff-AC0[42]}.

\begin{proof}
Let $d \equiv 2 \bmod 6$, let $f$ be a symmetric function on $n$ inputs, and let $g$ be its companion function; that is, for every $\mathbf{x}$, $f(\mathbf{x}) = g(|\mathbf{x}|_1)$. We begin with the same opening move as Oliveira, Santhanam, and Srinivasan~\cite{OliveiraSS19}, observing that \[\displaystyle f(\mathbf{x}) = \bigvee_{i \in g^{-1}(1)} \bigwedge_{j \ne i} D_{i,j},\] where $D_{i,j}(\mathbf{x}) = 1$ if $|\mathbf{x}|_1 = i$ and $D_{i,j}(\mathbf{x}) = 0$ if $|\mathbf{x}|_1 = j$ (and has otherwise arbitrary behavior). Thus it suffices to construct circuits $C_{i,j}$ computing functions consistent with $D_{i,j}$.

When $|i-j| \ge n^{7/13}$, Lemma \ref{lemming3} gives an $\mathsf{AC}^0$ circuit $C_{i,j}$ of depth $d-1$ and size $\exp({\tilde{O}(n^{6/(13(d-2))})})$ computing $D_{i,j}$.

When $|i-j| \le n^{7/13}$, we observe that a circuit for $\MOD_m$ suffices, with $m > n^{7/13}$. We take $\alpha = 7/13$ in Theorem \ref{tradeoff-MOD}.
Then we have a circuit $C'_{i,j}$ of depth $d-1$ and size $\exp({\tilde{O}(n^{6/(13(d-2))})})$ computing the $\text{MOD}_m$ function on $2n$ inputs, where $m > n^{7/13}$.
We now take $C_{i,j}(\mathbf{x}) = C'_{i,j}(\mathbf{x} 1^{m-i}0^{n-m+i})$.
Finally, we set \[\displaystyle C = \bigvee_{i \in g^{-1}(1)} \bigwedge_{j \ne i} C_{i,j}.\]
We can collapse the output $\text{AND}$ gates of all of the $C_{i,j}$ into the second layer $\text{AND}$ gates, so $C$ has depth $d$ and size $\exp({\tilde{O}(n^{6/(13(d-2))})})$, as desired.
\end{proof}

More generally, for certain $m$ which are the product of $r$ primes, we can improve the results of Theorem \ref{tradeoff-AC0[42]}. Recall from the introducion that we defined a product $m$ of primes $q_1,\ldots,q_r$ to be {\bf good} if every prime factor of $\phi(m)$ divides $m$, and we noted that the primorial $m = p_r\#$, the product of the first $r$ primes, is good.

\begin{reminder}{\Cref{thm:acc-symmetric}}
Let $m$ be a good product of $r$ primes. 
For every symmetric function $f$ on $n$ inputs and every depth $d \ge 4$ congruent to $1$ modulo $3$, there exists an $\mathsf{AC}^0[m]$ circuit of depth $d$ and size $\exp({\tilde{O}(n^{3/((r+3)(d-1)-3)})})$ computing $f$.
\end{reminder}

\begin{proof}
Let \[\displaystyle m = \prod_{a=1}^r p_a\] be a good product of $r$ primes.
For each $a \in [r]$, let $\displaystyle s_a = \left\lceil \alpha \log_{p_a} n \right\rceil$ for some $\alpha$ to be defined later.
By Lemma \ref{lemming2}, there are arithmetic circuits $C_{a,b}$ over $\mathbb{F}_{p_a}$ of depth $(2/3)(d-1)$
and size $\displaystyle n^{O(n^{3 \alpha / (d-1)})}$ computing the $p_a^b$-th elementary symmetric polynomial in $2n$ inputs over $\mathbb{F}_{p_a}$.
By Lemma~\ref{lemming5}, we can convert these into $\mathsf{AC}^0[m]$ circuits $C'_{a,b}$ of depth $d-1$ and size $\displaystyle 2^{\tilde{O}(n^{3 \alpha / (d-1)})}$.
When $\displaystyle |i-j| \le n^{\alpha r}$, $i \not \equiv j \bmod p_a^{s_a}$ for at least one $a$ by the Chinese Remainder Theorem, so we can construct a circuit $E_{i,j}$ computing $D_{i,j}$ by taking \[E_{i,j}(\mathbf{x}) = \neg C'_{a,b}(\mathbf{x}, 0^{p_a^{s_a}-j}1^{n+j-p_a^{s_a}})\] for some pair $(a,b)$.
When $\displaystyle |i-j| \ge n^{\alpha r}$, we use Lemma~\ref{lemming3} to get a circuit $E_{i,j}$ of depth $d-1$ and size $\exp({\tilde{O}(n^{(1-\alpha r)/(d-2)})})$ computing $D_{i,j}$.
All of the $E_{i,j}$ have $\text{AND}$ gates as output gates, so we take $\alpha = \frac{d-1}{(r+3)(d-1)-3}$ to balance the sizes of the two circuit constructions and complete the proof as per Theorem~\ref{tradeoff-AC0[42]}.
\end{proof}

It is worth noting that when $2 \mid m$, we can improve this construction slightly.
When $2 \mid m$ (and $6 \mid d-1$), the $(r+3)(d-1)-3$ in the denominator of the double exponent instead becomes $(r+\frac{7}{2})(d-1)-3$.

\section{Conclusion}

We believe our work demonstrates that $\CC^0$ circuits are not as weak as conventional wisdom anticipates, even at depth three. We hope that researchers seriously consider (possibly refuting) the SYM~$\circ$~AND hypothesis, as it stands in the way of obtaining significantly smaller $\CC^0$ and $\ACC^0$ circuits for symmetric functions. 

A natural next step would be to explore how much further our constructions can be pushed beyond symmetric functions. Our Theorem~\ref{thm:optimal} already demonstrates that $\TC^0$ circuits with linearly many gates and linear fan-in can be non-trivially simulated with $\CC^0$ circuits in subexponential size. Another question is whether $\NC^1$ circuits or Boolean formulas can be simulated similarly. For another example, it is well-known that time $t$ and space $s$ computations can be simulated with depth-three $\AC^0$ circuits of size $2^{O(\sqrt{t\cdot s})}$; this follows from efficient simulations in the polynomial hierarchy of space-bounded computation~\cite{Nepomnjascii70}. Could the size of this construction be improved, using $\MOD_m$ gates? If such an improved circuit could be constructed in a uniform way, it would likely imply new time-space lower bounds for decision problems in $\PP$ or the counting hierarchy~\cite{AllenderKRRV01}. However, even a non-uniform construction would be very interesting.

\paragraph*{Acknowledgements.} We thank Arkadev Chattopadhyay and Kristoffer Arnsfelt Hansen for useful pointers and discussion.

\bibliographystyle{alpha}
\bibliography{papers2}

\appendix 

\section{Proof of Theorem~\ref{thm:optimal}}
\label{sec:optimal}

Let us recall the SYM~$\circ$~AND Hypothesis and its consequence stated in the introduction.

\begin{reminder}{Hypothesis~\ref{hyp:sym-and}}
There are constants $c, k > 1$ such that for sufficiently large $n$, there is a function $f : \{0,1\}^n \rightarrow \{0,1\}$ computable by $\TC^0$ circuits of depth $c$ with at most $\tilde{O}(n)$ gates where each gate has fan-in $\tilde{O}(n)$, such that $f$ does not have an $\exp(O(n^{1/k}))$ size $\SYM \circ \AND$ circuit.
\end{reminder}

\begin{reminder}{Theorem~\ref{thm:optimal}}
Assuming the $\SYM \circ \AND$ Hypothesis (Hypothesis~\ref{hyp:sym-and}), 
there is a fixed $\alpha > 0$ such that for \emph{every} $m$ and $d$,
every depth-$d$ $\ACC^0[m]$ circuit computing the MAJORITY function on $n$ inputs requires size at least $\exp(n^{\frac{\alpha}{rd}})$ for sufficiently large $n$, where $r$ is the number of distinct prime factors of $m$.
\end{reminder}

We prove the contrapositive. We start with the negation of the theorem's conclusion:

\begin{quote}
Suppose for every $\alpha > 0$, there is some modulus $m$ which is a product of $r$ primes, along with some depth $d$, such that MAJORITY can be computed by a depth-$d$ $\ACC^0[m]$ circuit of size $\exp(O(n^{\frac{\alpha}{rd}}))$.
\end{quote}

Assuming the above, we will refute the $\SYM \circ \AND$ Hypothesis: we will show for all $c, k > 1$ and every function $f$ computable by the appropriate depth-$c$ $\TC^0$ circuits, $f$ has an $\exp(O(n^{1/k}))$ size $\SYM \circ \AND$ circuit. 

Let $c,k > 1$ be arbitrary. Let $C$ be a $\TC^0$ circuit $C$ with depth $c$ and $\tilde{O}(n)$ gates each of fan-in at most $\tilde{O}(n)$. Suppose we substitute each MAJORITY gate of $C$ with a copy of the assumed $\ACC^0[m]$ circuit. We obtain a $\ACC^0[m]$ circuit $C'$ of depth at most $c\cdot d$ and of size $\exp(\tilde{O}(n^{\frac{\alpha}{rd}}))$ such that $C'$ is equivalent to $C$.

Chen and Papakonstantinou~\cite{ChenP19} prove that for every depth-$d'$ size-$s$ circuit $D$ over AND, OR, and $\MOD_m$ gates, where $m$ is the product of $r$ distinct primes, $D$ is equivalent to a SYM~$\circ$~AND circuit $D'$ of size at most \[S'(s,m,r,d') = 2^{(m \log s)^{10rd'}}.\]

Applying their reduction to our $C'$, we obtain a $\SYM \circ \AND$ circuit $C''$ of size $\exp(\tilde{O}(n^{10 \alpha c}))$ that is equivalent to our original circuit $C$. For all $\alpha < 1/(10 c k)$, we obtain a $\SYM\circ \AND$ circuit equivalent to $C$ with size $\exp(O(n^{1/k}))$.

\end{document}